\documentclass[doublecol]{epl2}

\def\be{\begin{equation}}
\def\ee{\end{equation}}
\def\bea{\begin{eqnarray}}
\def\eea{\end{eqnarray}}

\title{Condensation phase transition in nonlinear fitness networks}

\author{Guifeng Su\inst{1} \and Xiaobing Zhang\inst{2} \and Yi Zhang\inst{3}}
\shortauthor{Guifeng Su  \etal}

\institute{
  \inst{1} Institute of Theoretical Physics and Department of Physics, East China Normal
  University - Shanghai 200241, People's Republic of China\\
  \inst{2} School of Physics, Nankai University - Tianjin 300371, People's Republic of China \\
  \inst{3} Department of Physics, Shanghai Normal University - Shanghai 200230, People's Republic of
China}


\abstract{We analyze the condensation phase transitions in
out-of-equilibrium complex networks in a unifying framework which
includes the nonlinear model and the fitness model as its
appropriate limits. We show a novel phase structure which depends on
both the fitness parameter and the nonlinear exponent. The
occurrence of the condensation phase transitions in the dynamical
evolution of the network is demonstrated by using
Bianconi-Barab\'asi method. We find that the nonlinear and the
fitness preferential attachment mechanisms play important roles in
formation of an interesting phase structure.}

\begin{document}

\maketitle

\section{Introduction}
\label{intro} Condensation phenomena emerge in various physical
contexts, to name a few, the well-known Bose-Einstein condensation
(BEC) in dilute atomic gases~\cite{Bose, BEC95, Bradley95}, jamming
in traffic flow~\cite{Evans96, Chowdhury00}, wealth condensation in
macroeconomi\-es~\cite{Burda02}, and condensation in zero-range
process (ZRP, see e.g., recent review~\cite{Evans05} and references
therein).
Since the pioneered research on complex networks~\cite{Watts98, BA,
BA00, Reviews, Reviews02, Reviews03, Books, Books04, Books03}, in
the last decade, condensation phenomena, i.e., condensation of links
(or edges) in complex networks has also been widely
discussed~\cite{Bouchaud00, Krapivsky00, Barabasi01, Burda01,
Godreche01, Bauer02, Berg02, Rodgers02, Doro03a, Doro05, Doro03b,
Farkas04, Noh05a, Noh05b, Ohkubo05a, Ohkubo05b, Ohkubo05c, Ohkubo07,
Godreche05}. In the context of complex networks,
the condensation phase corresponds to the situation that a single
node captures a macroscopic finite fraction of total links/edges. It
has been found that condensation phenomena can occur in both growing
and non-growing complex networks. The condensation phase transitions
occurring in non-growing networks~\cite{Burda01, Bauer02, Berg02,
Doro03a, Doro05, Doro03b, Farkas04} are formally equivalent to that
in balls-in-boxes model~\cite{Bialas97}, and hence has been well
studied to a large extent via methods of equilibrium statistical
mechanics. While for growing complex networks, the appearance of the
condensation phase transitions during the dynamical evolution of the
network are particularly interesting due to its out-of-equilibrium
characteristics.

The tasks in this paper are two folds: first, we merge two important models on
this regard, the growing network with nonlinear preferential attachment (we will
refer to ``nonlinear model''~\cite{Krapivsky00} from now on), and the fitness
model~\cite{fitness} into a unifying framework--the nonlinear fitness model. We
then argue that the condensation phase transition appearing in fitness model and
that in nonlinear model stems from different mechanisms; Second, particularly
interesting, we reveal a novel phase structure in the model and this may increase
our understanding on the non-equilibrium phase transitions in dynamical evolution
of complex networks.

The nonlinear model is defined as follows. At each time step $t$, the newly-added
node created $m$ directed links to ones of the earlier existing nodes with $k$-link,
according to a probability, say, $\Pi$, that is proportional to some ``connection
kernel'' $k^{\gamma}$, $\Pi \propto k^{\gamma}$. Here the exponent $\gamma \geq 0$
reflects the tendency of preferential linking to a popular node and hence controls
the preferential attachment. In Ref.~\cite{Krapivsky00}, P. L. Krapivsky {\it et al.}
had discussed the cases of different choices on exponent $\gamma$, i.e., $\gamma =
1$, $\gamma < 1$ and $\gamma > 1$, for growing complex networks with connection
probability,
\be
\Pi^s = \frac{k_i^{\gamma}}{\Sigma_j k_j^{\gamma}}.
\ee

They proved that the number of nodes with $k$ links, $N_k$, follows a power law
distribution in the case that $\gamma$ closes to unity. While in the case of
$\gamma < 1$, the distribution shows a stretched exponential form. For $\gamma > 1$,
that is so called the super-linear case, the model exhibits a condensation phase
transition. Especially when $\gamma > 2$, there exists a limiting situation where
the most connected node links to almost all the other nodes in the network, this
corresponds to a ``winner-takes-all'' phenomenon. In this case, the degree of the
most connected node follows $k_{max} \sim t$.

A similar condensation phase, or so-called ``winner-takes-all'' phase, also appears
in fitness model of the complex networks~\cite{fitness, Barabasi01, Rodgers02}. In
fitness model of growing networks, a {\it fitness} parameter, $\eta_i$, which
represents an internal superiority of the $i$-th node, is introduced and chosen
randomly from some distribution. As a result, the well-known Barab\'asi-Albert
scale-free network~\cite{BA} is generalized to the Bianconi-Barab\'asi (B-B) fitness
model of complex network. One may assign the $i$-th node the fitness $\eta_i$
according to its ``energy level'' $\varepsilon_i$ which satisfy some distribution
$g(\varepsilon)$ through the relation
\be
\eta_i = e^{-\beta \varepsilon_i},
\label{etai}
\ee
where $\beta$ can be identified as inverse temperature, i.e., $\beta = 1/T$. In this
growing complex network with fitness, the connection probability $\Pi^{f}$ that a new
node connects one of its $m$ links to an existing node $i$ at each time step $t$ is
defined by
\be
\Pi^{f} = \frac{\eta_i k_i}{\sum_j \eta_j k_j},
\label{Pfi}
\ee
where $k_i$ is the degree (the number of links occupied by one node) of node
$i$. One may introduce a partition function $Z_{t}$ as
\be
Z_{t} = \sum_{j = 1}^t e^{-\beta \varepsilon_j} k_j(\varepsilon_j, t, t_j).
\ee

This model can be solved in a mean-field approximation and the condensation phase
transition process is described by the ``chemical potential'' $\mu$ formally defined
as follows,
\be
e^{-\beta \mu} = \lim_{t \rightarrow \infty} \frac{\overline{Z_{t}}}{mt},
\label{muf}
\ee
where $\overline{Z_{t}}$ is the partition function $Z_t$ averaged over some normalized
distribution $g(\varepsilon)$~\cite{Barabasi01}.

The chemical potential $\mu$ is determined by the self-consistent equation
\be
I(\beta, \mu) = \int d \varepsilon g(\varepsilon) n(\varepsilon) = 1 ,
\label{selfc}
\ee
where $n(\varepsilon)$ is the occupation number, i.e., the number of
links attached by the preferential attachment mechanism to nodes with
``energy'' $\varepsilon$. Very interestingly, it was proved~\cite{Barabasi01}
that $n(\varepsilon) = 1/(e^{\beta(\varepsilon - \mu)} - 1)$, this
is nothing but Bose-Einstein (BE) statistics. When $\mu < 0$, the
network is in so-called ``fit-get-rich'' (FGR) phase. While in the
thermodynamic limit $t \rightarrow \infty$ if $I(\beta, 0) < 1$,
i.e., the self-consistent equation (\ref{selfc}) has no solution,
and hence a BEC phase transition occurs. There exists a critical
temperature $T_C = 1/\beta_C$ such that $I(\beta, 0) < 1$ for $T <
T_C$. This condensation phase transition was demonstrated as well by
numerical simulations in Ref.~\cite{Barabasi01}.

Both the fitness model and nonlinear model of the complex network
experience the condensation phase transition during their dynamical
evolutions, two questions are naturally arisen: whether or not the
condensation phase transitions in these models have some common
underlying relationship~\cite{Albert02, Goltsev08}, and the role
that different preferential attachment mechanism in either model
plays during the phase transition of the network. In order to answer
such questions, and as well, to explore the possible phase structure
in out-of-equilibrium evolution in complex networks, we propose the
nonlinear fitness model of complex networks which includes the
nonlinear model and the fitness model as its appropriate limits,
as we will show in the following.

\section{The nonlinear fitness growing network}
\label{sec:1}

It is clear that the preferential attachment mechanism controls the network
topology in the nonlinear network and the fitness parameter $\eta_i$ does in
the network with fitness. To bridge the divergence between the two models and
to give a general description of the phase structure, we employ a new connection
probability, $\Pi^{sf}_i$, as follows,
\be
\Pi^{sf} = \frac{\eta_i k^{\gamma}_i}{\sum_j \eta_j k^{\gamma}_j}.
\label{Pi1}
\ee
This new probability includes the original connection probability $\Pi^{s}$ in
the nonlinear model and $\Pi^{f}$ in the fitness model, respectively, as its
proper limits: in the limit $\gamma \rightarrow 1$ one goes back to the
preferential attachment probability of the fitness model, and in the limit
$\eta \rightarrow 1$, one recovers the nonlinear model of growing network.
This attachment mechanism now addresses the dynamical evolution of the network.
In order to identify the condensation phase transition of this nonlinear fitness
model of network, similar to the Bianconi-Barab\'asi method, (see Eqn. (\ref{muf})),
we formally define the chemical potential $\mu$~\cite{Barabasi01},
\be
\mu = - \frac{1}{\beta} \lim_{t \rightarrow \infty} \ln \frac{\overline{Z^c_t}}{mt}.
\label{mu}
\ee
where, $\beta$ is again the inverse temperature, $m$ is the number of newly added
links at each time step $t$, and
$\overline{Z^c_t}=\sum_{j = 1}^t e^{-\beta \varepsilon_j} k^\gamma (\varepsilon_j, t, t_j)$
is the partition function in our nonlinear fitness model, $t \rightarrow \infty$ plays
the role of thermodynamic limit.

The corresponding rate equation is,
\be
\frac{\partial k_i(\varepsilon_i, t, t_i)}{\partial t} = \frac{e^{-\beta
\varepsilon_i} k_i^{\gamma} (\varepsilon_i, t, t_i)}{\overline{Z^c_t}}.
\ee

Note that in $\gamma \rightarrow 1$ limit, the rate equation of B-B fitness model
is restored and one naturally expects a BEC phase transition at low enough
temperature~\cite{Barabasi01}. However, in general, the chemical potential
$\mu (\gamma, T)$ is a function of both temperature $T$ (or the fitness $\eta$)
and the nonlinear exponent $\gamma$, i.e., the phase structure of this nonlinear
fitness network is controlled by these two parameters, instead only one parameter
in either model ($T$ in the original B-B fitness model or $\gamma$ in the nonlinear
one). These parameters both affect the phase transition during the dynamic evolution
of the network. This fact leads to a more complicated behavior of $\mu$ in current
model than that in the fitness or the nonlinear model alone. Here we apply the method
of rate equation of {\it degree}, used in the B-B fitness model~\cite{Barabasi01},
rather than that of {\it connectivity distribution}, used in the nonlinear
model~\cite{Krapivsky00}. This is due to the fact that the fitness parameter $\eta_i$
is different for each node $i$, i.e., $\eta$ is a local, not a global parameter. This
suggests that the fitness parameter $\eta$ and the exponent parameter $\gamma$ which
respectively controls the preferential attachment mechanism are different in nature --
$\eta_i$ reflects an ``inner'' property of the $i$th node, while $\gamma$ controls the
global evolution of the network. This important difference between the B-B fitness
model and the nonlinear model leads to the different dynamical evolution result of the
complex networks.

\section{Numerical simulations and discussions}
\label{sec:2}

Similar to the original B-B fitness model, we identify the non-condensation-condensation
phase transition by the change of sign of the chemical potential $\mu (\gamma, T)$, i.e.,
when $\mu$ experiences a change from negative value to positive value within some regime,
that change implies the critical point of the corresponding phase transition. In addition,
in principle, the phase transition occurs in the thermodynamic limits $t \rightarrow \infty$.
However, in general situations, taking such thermodynamical limit is not realistic, one has
to resort to the numerical simulations. We numerically compute the chemical potential $\mu$
according to Eqn.(\ref{mu}).

\begin{figure}
\includegraphics[width=8cm]{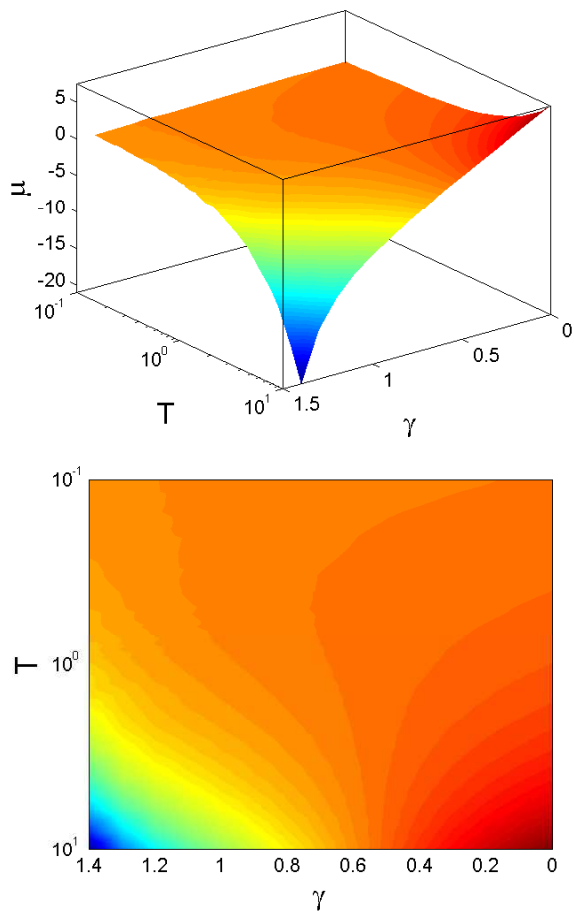}\\
\caption{(Color online) Three dimensional illustration of the phase structure
in the nonlinear fitness complex network (Top), and its projection on the
$\gamma - T$ plane (Bottom). In top panel, it shows the chemical potential
$\mu$ as a function of temperature $T$ and the nonlinear exponent $\gamma$,
as defined in Eqn. (\ref{mu}). In this simulation we take the number of links
per node (per time step) $m = 2$, and energy level distribution
$g(\varepsilon) = 2 \varepsilon$ (normalized), and the total number of time
steps $t = 10^3$, average over $100$ runs. \label{fig1}}
\end{figure}

Throughout our simulations we fix the number of links per node (per time step) $m = 2$. For
simplicity, we take energy level distribution $g(\varepsilon) = C \varepsilon$ with
normalization constant $C = 2$, and the total number of time steps $t = 10^3$, average over
$100$ runs. The main numerical simulation results are plotted in a three dimensional (3d)
figure (see Fig.~\ref{fig1}), in which three axes of the figure are exponent $\gamma$,
temperature $T$ and chemical potential $\mu$, respectively. In this ``phase diagram", one
can see that the role that the temperature $T$ or exponent $\gamma$ plays in the formation
of such a phase structure during the dynamical evolution of the network. For instance, for
$\gamma = 1$, when one lowers the temperature, a transition from FGR phase (high temperature
phase) to BEC phase (low temperature phase) occurs~\cite{Barabasi01} as what we expect.
Similarly, if one fixes the temperature, say $T = 2.0$, a phase transition would also occur
when the exponent $\gamma$ is varied (between $\gamma = 0$ and $\gamma = 1.5$ in
Fig.~\ref{fig1}). However, one should keep in mind that the current phase transition is
different with that in nonlinear model because no fitness (temperature) effect was taken into
consideration in the latter model. On the other hand, in both cases, the signal of the phase
transition lies on the change of sign of the chemical potential. Note that we do not take
absolute value of the chemical potential, which is different with Ref.~\cite{Barabasi01}.
We see that in Fig.~\ref{fig1} the chemical potential $\mu$ experiences a change from
positive values to negative ones when the exponent $\gamma$ and/or the temperature $T$ vary
in some regime. Now the whole phase structure is richer than that in the B-B fitness model
and in the nonlinear model, respectively, e.g., the BEC phase and the FGR phase in the B-B
fitness model now are only parts of the new phase diagram in $\gamma = 1$ limit.

\begin{figure}
\includegraphics[width=8cm]{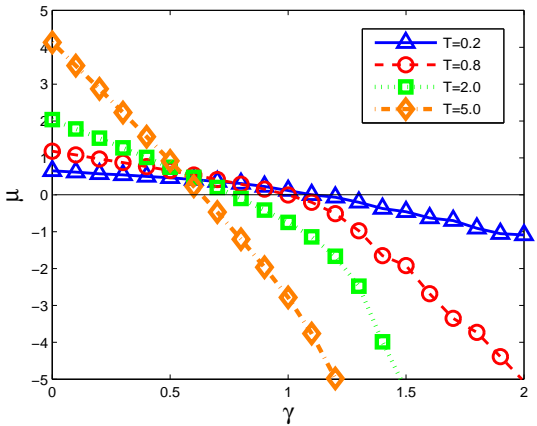}\\
\caption{(Color online) The chemical potential $\mu (\gamma, T)$ as a function of
the exponent $\gamma$ but with temperature fixed, in a linear-linear
scale. The four curves shown in the panel correspond to $T = 0.2$
(open triangles, solid line), $T = 0.8$ (open circles, dashed line),
$T = 2.0$ (open squares, dotted line), $T = 5.0$ (open diamonds,
dot-dashed line), respectively. All symbols are connected by lines
for eye guidance. \label{fig2}}
\end{figure}

\begin{figure}
\includegraphics[width=8cm]{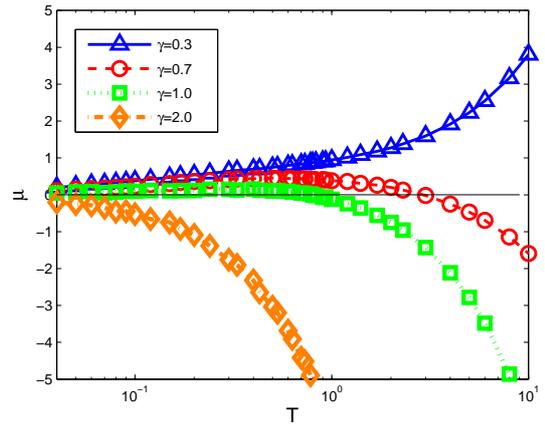}\\
\caption{(Color online) Similar to Fig.~\ref{fig2}, the chemical potential $\mu
(\gamma, T)$ as a function of temperature but with exponent $\gamma$
fixed, in a linear-log scale. The four curves shown in the panel
correspond to $\gamma = 0.3$ (open triangles, solid line), $\gamma =
0.7$ (open circles, dashed line), $\gamma = 1.0$ (open squares,
dotted line), $\gamma = 2.0$ (open diamonds, dot-dashed line),
respectively. Note that for $\gamma = 0.3$, the values of $\mu$-s
are all positive, while for $\gamma = 2.0$, the values of $\mu$-s
are all negative. \label{fig3}}
\end{figure}

The situation is more clear when we plot the chemical potential $\mu$ as a
function of exponent $\gamma$ in Fig.~\ref{fig2}, and of temperature $T$ in
Fig.~\ref{fig3}, respectively. These figures display the different ``cross
section" views of 3d ``phase diagram" of Fig.~\ref{fig1}. Some new features of
the phase transition can be read out from these plots. In Fig.~\ref{fig2}, we
show the chemical potential $\mu$ versus the exponent $\gamma$ with temperature
$T$ fixed. It can be seen from the figure that as temperature increases, the
critical value of exponent for phase transition $\gamma_C$ decreases, e.g., for
$T = 0.2$, $\gamma_C \sim 1.2$, while for $T = 5.0$, $\gamma_C \sim 0.6$, roughly.
This tells us that the transition of phase occurs with weaker tendency of
preferential linking at relatively higher temperature.

\begin{figure}
\includegraphics[width=8cm]{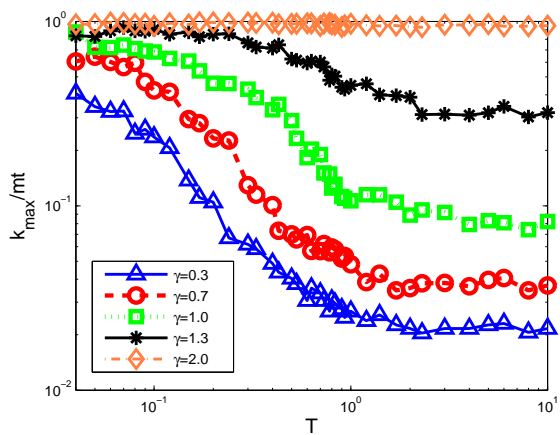}\\
\caption{(Color online) The occupation ratios of the most connected node,
$k_{max}/mt$, plotted as a function of temperature $T$, and of
different nonlinear exponent $\gamma$. The five curves in the panel,
from bottom to top, represent $\gamma = 0.3$ (open triangles, solid
line), $\gamma = 0.7$ (open circles, dashed line), $\gamma = 1$
(open squares, dotted line), $\gamma = 1.3$ (stars, solid line),
$\gamma = 2$ (open diamonds, dot-dashed line), respectively. Note
that the data points with $\gamma = 1$ are consistent with B-B's
results (see the corresponding curve of Fig.~\ref{fig3}. in
Ref.~\cite{Barabasi01}). \label{fig4}}
\end{figure}

In Fig.~\ref{fig3} we plot the chemical potential $\mu$ versus the temperature
$T$ with exponent $\gamma$ fixed in a linear-log scale. For data points with
$\gamma = 0.7$ (open circles, dashed line) and $\gamma = 1$ (open squares, dotted
line), the condensation phase transition is very similar to what happens in
Fig.~\ref{fig2}: the critical temperature $T_C$ of the phase transition decreases
as the exponent $\gamma$ increases, e.g., for $\gamma = 0.7$, $T_C \sim 3.0$, while
for $\gamma = 1$, $T_C \sim 0.8$. The latter case recovers the results of B-B BEC
phase transition, as it should. Fig.~\ref{fig3} also displays some new, interesting
aspects: the values of the chemical potential with $\gamma = 0.3$ (open triangles,
solid line) are all positive and those with $\gamma = 2$ are all negative. This
implies that the phase transition disappears eventually as $\gamma$ reaches some
specific critical value, and hence results in a single phase structure -- either a
FGR phase (with negative chemical potential), or a condensation phase (with positive
chemical potential). However, one can not always use this as an identification of
transition between the condensation phase and the non-condensation phase, this
point is illustrated in Fig.~\ref{fig4}.

We plot the occupation ratio, $k_{max}/mt$, of {\it the most connected} node, in
Fig.~\ref{fig4}. At the first sight from Fig.~\ref{fig4}, it seems that the network
is in a condensation phase at low enough temperature $T < T_C$ ($T_C$ is the critical
temperature), and in a non-condensation phase at high temperature, $T > T_C$. But note
that as the exponent $\gamma$ increases, the condensation phase appears in even
higher temperature region. Especially when $\gamma = 2$, the network enters a
condensation phase regardless of the value of temperature. This supports the
conclusion of nonlinear growing network in Ref.~\cite{Krapivsky00}. However, in addition
to the appearance of the condensation phase in original B-B fitness model and the
nonlinear model with $\gamma > 1$, we conclude from Fig.~\ref{fig4} that they are still
{\it different} phases because the chemical potential is positive ($\mu > 0$) for the
B-B fitness model at $T < T_C$, while negative ($\mu < 0$) for the nonlinear model. The
competitive relationship between $e^{-\beta \varepsilon_i}$ and $k^{\gamma}$ in partition
function play important roles in the dynamical evolution of corresponding complex networks
and hence the formation of such a novel phase structure. When the former factor
$e^{-\beta \varepsilon_i}$ is dominant, the network experiences a condensation phase
transition if $T < T_C$, and its chemical potential is positive, $\mu > 0$; while the
prevailing role of the factor $k^{\gamma}$ leads to the opposite side: the chemical
potential of the condensation phase is always negative, $\mu < 0$. It is also clear that,
from above figure, for $\mu > 0$ and $\gamma = 1$, the network with fitness preferential
attachment is condensed on the node with the {\it lowest} energy level (or the fittest
node) only if $T < T_C$ in the thermodynamic limit; for $\mu < 0$, the network experiences
a ``winner takes all'' phenomenon when $\gamma$ is large enough (e.g., $\gamma > 1.7$).
Due to aforementioned competitive relationship between two factors $e^{-\beta \varepsilon_i}$
and $k^{\gamma}$, the chemical potential of the non-condensation phase is changed, for
instance, see the curve with $\gamma = 0.3$ in Fig.~\ref{fig3}, though one has an overall
positive $\mu$ at high temperature, no condensation phase appears. The different phase
structures identified by the chemical potential $\mu$ stem from different dominant factors
of preferential attachment mechanism, they are distinct in nature.
In addition, as is well known, a significantly different manifestation of BEC
and FGR phases is that the fraction of the total number of links connected to the most
connected node tends to zero in thermodynamic limit for the latter and tends to a finite
value for the former. In current combined model, Fig.~\ref{fig4} shows that there exists a
similar situation for $\mu>0$ when the factor $\eta$ dominates the attachment mechanism.
while when the factor $\gamma$ dominates the attachment, the ratio tends to zero in
thermodynamic limit for $\mu<0$, which displays a FGR phase. However, note that if $\gamma$
is large enough, this occupation ratio can also approach to a finite value and a
condensation phase appears even for $\mu<0$.

\begin{figure}
\includegraphics[width=8cm]{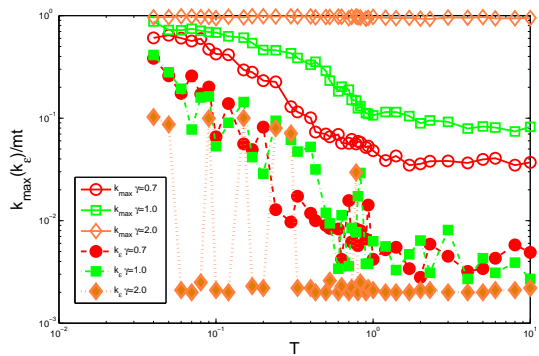}\\
\caption{(Color online) The occupation ratios of the most connected node,
$k_{max}/mt$, and that of the node with the lowest energy level,
$k_{\varepsilon}/mt$, plotted as a function of temperature $T$, and
of different nonlinear exponent $\gamma$'s. Data with hollow symbols
and solid lines correspond to $k_{max}/mt$, represent $\gamma = 0.7$
(open circles), $\gamma = 1$ (open squares), $\gamma = 2$ (open
diamonds), respectively. Data with solid symbols and non-solid lines
correspond to $k_{\varepsilon}/mt$, represent $\gamma = 0.7$ (solid
circles, dashed line), $\gamma = 1$ (solid squares, dot-dashed
line), $\gamma = 2$ (solid diamonds, dotted line), respectively.
\label{fig5}}
\end{figure}

A major difference between the condensations of fitness model and non-linear model
is that the condensation in the former occurs on the node with the lowest energy level (or
the highest fitness), instead in the latter model it only occurs on the first node of the
network. However, topologically it is hardly able to tell whether or not a graph has reached
condensation due to the non-linear preferential attachment or the BEC due to the fitness,
since in both condensation phases, the important topological parameter, i.e., the cluster
coefficient (CC) approaches to the same limit. If one trace the node with the lowest energy
level, then this node is more likely to be the {\it most connected} node in low temperature
region, since with large parameter $\beta=1/T$, the fitness parameter $\eta$ dominates the
dynamical evolution in combined attachment mechanism. As temperature increases, the
probability of the node with the lowest energy level being the most connected node is getting
smaller due to the impact of fitness $\eta$, as a result, in high temperature region, the
factor $k^{\gamma}$ dominates the attachment. This is shown in Fig.~\ref{fig5}, in which the
occupation ratios of the node with the lowest energy level, $k_{\varepsilon}/mt$, is plotted;
as well as that of the most connected node, $k_{max}/mt$, as function of temperature.
The former exhibits evidently larger fluctuations compared to the occupation ratio
of the most connected node $k_{max}/mt$, since the node with the lowest energy level is not
necessary the most connected node. As a result, in high temperature region, in which
$k^{\gamma}$ dominates the attachment ($\eta \rightarrow 1$), fluctuations are relatively
small, since as a global parameter, $\gamma$ does not depend on the local fitness
of a node, such that the earliest presented node could be the most connected node for large
$\gamma$, regardless the local fitness for $T>T_C$ ($T_C$ is the critical temperature); in
low temperature region, ($T<T_C$), fluctuations of the occupation ratio of a node with the
lowest energy level suggests that the emergence of condensation comes from the competition
of two sides -- the local fittest node (with fitness attachment) and the earliest presented
node (with super-linear preferential attachment), and this competition between makes
fluctuations larger and larger (e.g., see $\gamma = 2$ in Fig.~\ref{fig5}). This demonstrates
how both attachment factors $\eta$ and $\gamma$ control the evolution of the network
simultaneously.

\section{Conclusions}
\label{sec:3}

In conclusion, in this paper we analyze the condensation phase transitions in a unifying
framework which includes both the nonlinear model and the fitness model as its appropriate
limit. The goodness of this new framework is to, on one hand, bridge the differences
between the nonlinear model and the fitness model; On the other hand, allow us to identify
the different roles played by different factors in current model. Under appropriate limits
(e.g., $\gamma \rightarrow 1$ or $\eta \rightarrow 1$), the typical phase structures of the
original B-B fitness model and the nonlinear model are recovered respectively.

Through the numerical simulations of this nonlinear fitness model, we show a novel 3d phase
diagram. We employ the Bianconi--Barab\`asi method and numerically compute the chemical
potential to identify the critical points ($T_C$ and $\gamma_C$) of the phase transitions
of the networks. Our results and analysis directly answer the questions that asked in the
beginning, i.e., the condensation phase transitions in both the B-B fitness model and the
nonlinear model are in fact distinct in nature. As well, the preferential attachment
mechanism in our nonlinear fitness model depends on two factors: $k^{\gamma}$ and
$e^{-\beta \varepsilon_i}$ (or $\eta_i$). Both factors affect the phase structure of the
network and hence the network topologies during the evolution of the network. We reveal
that the competitive relationship between these two factors leads to the condensation phase
transitions, and hence transitions between different network topologies. We hope the current
study may increase our understanding to the non-equilibrium critical phenomena, particularly
the condensation phase transitions in the evolution of complex networks.

\acknowledgments We acknowledge to the Initiative Plan of Shanghai Education Committee
(Project No. 10YZ76) and the Scientific Research Foundation for the Returned Overseas
Chinese Scholars, State Education Ministry (SRF for ROCS, SEM) for their support.

\end{document}